\documentclass[aps,reprint,superscriptaddress,longbibliography]{revtex4-1}
\usepackage{}
\usepackage{amssymb}
\usepackage{amsmath}
\usepackage{dcolumn}
\usepackage{bm}
\usepackage{graphicx}
\usepackage{mathrsfs}
\usepackage[colorlinks,linkcolor=blue,anchorcolor=blue,citecolor=blue,urlcolor=black]%
{hyperref}

\begin{document}
\title{ Proposal for  a quantum delayed-choice experiment with  a spin-mechanical setup}
\author{Peng-Bo  Li}
\email{lipengbo@mail.xjtu.edu.cn}
\homepage{http://lipengbo.gr.xjtu.edu.cn/ }
\affiliation {Shaanxi Province Key Laboratory of Quantum Information and Quantum Optoelectronic Devices,\\
Department of Applied Physics, Xi'an Jiaotong University, Xi'an
710049, China}
\affiliation {Kavli Institute for Theoretical Physics China, CAS, Beijing 100190, China }

\author{Fu-Li  Li}

\affiliation {Shaanxi Province Key Laboratory of Quantum Information and Quantum Optoelectronic Devices,\\
Department of Applied Physics, Xi'an Jiaotong University, Xi'an
710049, China}

\begin{abstract}
   We describe an experimentally feasible protocol for performing a variant of the quantum delayed-choice experiment with massive objects.
   In this scheme, a single nitrogen-vacancy (NV) center in diamond driven by microwave fields is dispersively coupled to a massive mechanical resonator.
   A double-pulse  Ramsey interferometer can be implemented with the spin-mechanical setup, where the second Ramsey microwave pulse
   drives the spin conditioned on the number states of the resonator. The probability for finding the NV center in definite spin states exhibits interference fringes when the mechanical resonator is prepared in a specific number state. On the other hand, the interference is destroyed if the mechanical resonator stays in some other number states. The wavelike  and particlelike  behavior of the NV spin can be superposed by preparing the mechanical resonator in a  superposition of two distinct number states. Thus a quantum version of Wheeler's delayed-choice experiment could be implemented, allowing of  fundamental tests of quantum mechanics on a macroscopic scale.
\end{abstract}

\maketitle

\section{introduction}
Quantum mechanics predicts many counterintuitive behaviors for small objects. For instance, a single particle, such as a photon or an electron, can be in
several places at the same time. A quantum object can behave either as a particle or a wave---the particle-wave duality, which is at the heart of quantum mechanics.
To account for the weird behavior of quantum objects, Bohr introduced the principle of complementarity \cite{Bohr,nature-375-367}: either wave or
particle behavior to be observed depends on the
kind of experimental apparatus with which the quantum object is
measured. Hence, these two  incompatible aspects can never be observed simultaneously. This is well demonstrated by sending a single photon into a Mach-Zehnder (MZ) interferometer with  two
detectors placed at the two outputs \cite{QO}. A photon, split by the first beam splitter, travels along two paths with a tunable phase difference, and is finally recombined (or not) at the second beam splitter before detection. If the second beam splitter is present, we can observe interference patten, representing  wavelike behavior. On the other hand, if the second beam splitter is absent, the  photon's path can be known and only a click with probability $\frac{1}{2}$ in one of the two detectors occurs, showing particle
properties.

One can conclude that these two different experimental configurations---the second beam splitter present or absent---give different experimental outcomes.  It seems that the photon may know in advance the type of detecting device, via
a hidden variable, and could thus decide which
behavior to exhibit. To examine this idea, Wheeler formulated  the delayed-choice experiment \cite{Wheeler-1,Wheeler-2,Leggett, RMP-88-015005}. In this gedanken experiment, the choice whether to insert the second beam splitter is delayed with respect to
the photon entering the interferometer. The choice of inserting or removing the second beam splitter  is
classically controlled by a random number generator. Thus, the photon could not have
known in advance  the kind of experiment with which it
will be confronted and which behavior it should
exhibit. Wheeler's thought experiment has
been implemented experimentally with various systems \cite{ RMP-88-015005,pra-35-2532,pra-54-5042,prl-84-1,SCI-315-966}.

Recently,  a quantum version of the delayed-choice experiment has been
proposed \cite{ prl-107-230406}, where a quantum ancilla is employed to coherently control the
second beam splitter of the interferometer. By this,  the second beam splitter can be in a
superposition of being present and absent, and consequently the photon must be in a
superposition of particle and wave at the same time. Contrary to Bohr's opinion, one does not need to change the
experimental setup in order to measure complementary
properties (wave and particle). Quantum delayed-choice experiments have been performed with nuclear magnetic resonance \cite{pra-85-022109, pra-85-032121}, optical \cite{SCI-338-634,SCI-338-637,nphton-6-600}, and superconducting circuit \cite{prl-115-260403} systems, all of which however test quantum mechanics on the microscopic level. It will be of great interest to test quantum mechanics on a  macroscopic scale, particularly with respect to
counterintuitive effects induced by the particle-wave duality. This is also particularly relevant to fundamental studies of the quantum-to-classical crossover,
as one moves from the microscopic to the macroscopic world.

In recent years significant advances in the control of nanoscale mechanical resonators have been achieved  \cite{PR-511,nature-464-697,nature-475-359,nature-478-89,prl-116-147202}, which culminated in the cooling of mechanical oscillators down to the
ground state \cite{nature-464-697,nature-475-359,nature-478-89}. An attractive
route in this field now is to couple single electronic spins to mechanical resonators and thereby form hybrid spin-mechanical systems, which have been extensively investigated both for fundamental research and practical applications \cite{prb-79-041302,natphys-6-602,np-7-879, SCI-335-1603,nl-12,Natcomm-6-8603,prl-110-156402,prl-111-227602,prl-113-020503,Natcomm-5-4429,prappl-5-034010,prl-117-015502,prl-116-143602,prl-112-190402,prappl-4-044003,pra-80-022335,pra-81-042323,pra-88-033614}.
It is thus very appealing to perform the quantum delayed-choice experiment with
the spin-mechanical system. This will test the most fundamental aspects of quantum mechanics on a macroscopic scale, and help us to deeply understand quantum mechanics.

Here, we suggest an experimentally feasible protocol for performing a variant of the quantum delayed-choice experiment in a spin-mechanical system.
In this scheme, a single NV center \cite{PR-528-1}, driven by microwave fields,  is dispersively coupled to a  mechanical resonator, which enables spin rotations conditional
on the quantum state of the  resonator. We propose to implement the quantum delayed-choice experiment by using a Ramsey interferometer \cite{OC-173-265,nature-411-166}, in which
microwaves act as beam splitters for the spin states of the NV center. In this proposal, the mechanical motion controls the second spin rotation, enabling it
in a superposition of being active and inactive.  We find that, the probability
for finding the NV center in definite spin states exhibits interference patten if the mechanical
resonator is prepared in a specific number state. On the other hand, the interference is destroyed
if the mechanical resonator is in some other number states. Therefore, the wavelike and particlelike behavior of the NV spin can be superposed by preparing the mechanical resonator in a superposition of two distinct number states. This provides an alternative  way to perform the quantum  delayed-choice experiment, allowing us to test quantum mechanics on a macroscopic scale.

\section{the proposed quantum delayed-choice experiment}
\begin{figure}[t]
\centerline{\includegraphics[totalheight=2.5in,clip]{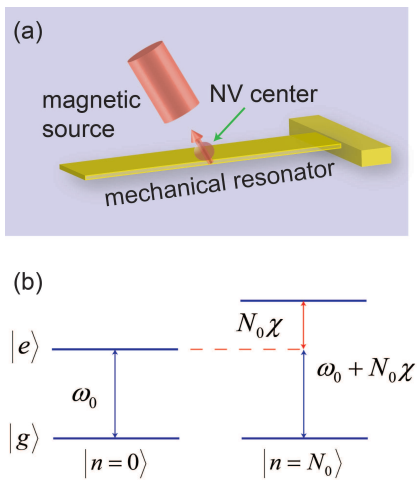}}
\caption{(Color online) (a) Schematic of a single NV center in a diamond resonator  dispersively coupled to the  mechanical motion.  (b) Energy-level diagram of the spin-oscillator system. In the rotating frame of the resonator, the states $\vert g,n\rangle,n=0,1,...$ have the same energy.}
\end{figure}
We consider a spin-mechanical setup depicted in Fig.1 (a), where a mechanical resonator with oscillation
frequency $\omega_m$ is dispersively coupled to a single NV center with ground state $\vert g\rangle$ and excited state
$\vert e\rangle$.  NV centers can couple to a mechanical resonator  either through
magnetic coupling \cite{prb-79-041302,natphys-6-602,np-7-879, SCI-335-1603,nl-12,Natcomm-6-8603,prl-117-015502} or strain-induced coupling \cite{prl-110-156402,prl-111-227602,prl-113-020503,Natcomm-5-4429,prappl-5-034010,prl-116-143602}.
The dispersive spin-motion coupling
is described by the following Hamiltonian ($\hbar=1$)
\begin{eqnarray}
\hat{\mathcal {H}}_{s}&=& \omega_0\vert e\rangle\langle e\vert+\omega_m \hat{a}^\dag\hat{a} +\chi  \hat{a}^\dag\hat{a}\vert e\rangle\langle e\vert,
\end{eqnarray}
where $\omega_0$ is the transition frequency between $\vert g\rangle$ and $\vert e\rangle$, $\hat{a}$ is the destruction operator for
the mechanical mode, and $\chi$ is the dispersive coupling strength.
The eigenstates of $\hat{\mathcal {H}}_{s}$ are $\vert g,n\rangle$  and $\vert e,n\rangle$ with a resonator
excitation number $n=0,1, ...$. As shown in Fig.1 (b), the difference of the energy
shifts of level $\vert g,n\rangle$ and $\vert e,n\rangle$, i.e., $\Delta^n_{eg}=\omega_0+n\chi$, which depends explicitly on the
number $n$ of phonons, will determine the
effective resonance frequency of the  $\vert g\rangle\leftrightarrow\vert e\rangle$ transition.
In this case, transitions inside a chosen subspace may be
tuned to resonance, while other transitions remain out of resonance, producing a selective drive in the spin-motion
Hilbert space \cite{prl-87-093601,prl-105-050501,pra-87-042315,pra-92-040303}. Once this frequency adjustment is made for
one specific subspace $\{\vert g, N_0\rangle, \vert e,N_0\rangle\}$, spin rotations can be realized conditional
on the number states of the resonator.

\begin{figure}[b]
\centerline{\includegraphics[bb=136 261 600 648,totalheight=2.3in,clip]{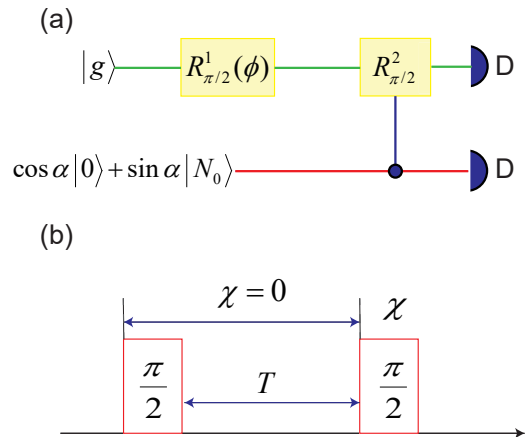}}
\caption{(Color online) (a) Schematic of the quantum delayed-choice experiment where the mechanical motion controls the second Ramsey rotation.
(b) Sequence of microwave pulses and related procedures. }
\end{figure}
We now consider the implementation of a Ramsey interferometer, where the second Ramsey sequence induces
spin rotations conditioned on the number states of the resonator. As shown in Fig. 2, we describe this Ramsey setup in two main steps.
In step (i), the spin is initially prepared in state $\vert g\rangle$, and we apply a $\pi/2$ microwave pulse with Rabi frequency $\Omega_1(t)$.
In this process, the
spin-motion coupling needs to be switched off, which can be made by keeping the spin transition far out of resonance with
the resonator mode.  In this case, the spin is rotated into a superposition state
\begin{eqnarray}
\vert \psi_p\rangle&=& R^1_{\pi/2}(\phi)\vert g\rangle\nonumber\\
&=&\frac{1}{\sqrt{2}}(\vert g\rangle+ie^{-i\phi}\vert e\rangle).
\end{eqnarray}
In order to tune the quantum phase difference  $\phi$, we subject the spin to a pulse of magnetic field for a time $T$, which shifts the spin
transition frequency with a variable detuning $\Delta$ and thus $\phi$ by a variable amount.

\begin{figure}[b]
\centerline{\includegraphics[bb=27 220 251 376,totalheight=1.8in,clip]{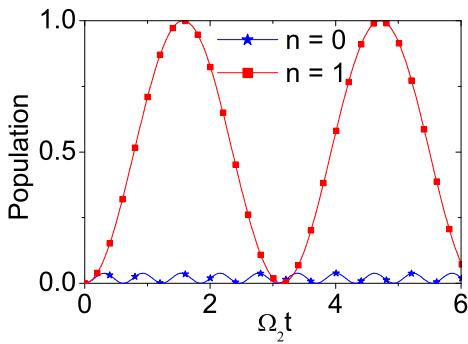}}
\caption{(Color online) Selected rotations conditioned on the number state of the resonator. Here the population of the state $\vert e\rangle$ is displayed.
The related parameters are $\chi=10\Omega_2, N_0=1$, and the the initial spin state is $\vert g\rangle$. }
\end{figure}

In step (ii), we turn on the dispersive coupling between the spin and the mechanical resonator, and achieve an unitary rotation $R^2_{\pi/2}(0)$ between the selected levels
$\{\vert g, N_0\rangle, \vert e,N_0\rangle\}$ by applying another $\pi/2$ pulse with the driving frequency $\omega_0+N_0\chi$. The corresponding  Rabi frequency $\Omega_2(t)$ should be much smaller than the dispersive coupling strength $\chi$ to ensure the rest of the system will not be affected by this drive, i.e.,
$\vert \Omega_2\vert\ll \chi$. If the resonator is prepared in the phonon state $\vert N_0\rangle$, the rotation $R^2_{\pi/2}(0)$ performs the following
transformations $\vert g\rangle \rightarrow (\vert g\rangle+i\vert e\rangle)/\sqrt{2}$ and $\vert e\rangle \rightarrow (\vert e\rangle+i\vert g\rangle)/\sqrt{2}$. This procedure results in the following state
\begin{eqnarray}
\vert \psi_w\rangle&=& R^2_{\pi/2}\vert \psi_p\rangle\nonumber\\
&=&-i[\sin (\phi/2)e^{i\phi/2}\vert g\rangle-e^{-i\phi/2}\cos (\phi/2)\vert e\rangle].\nonumber\\
\end{eqnarray}
However, if the mechanical resonator is instead in some other phonon states such as the ground state $\vert 0\rangle$, the rotation $R^2_{\pi/2}(0)$ will not occur and the spin remains in its original state $\vert \psi_p\rangle$. The above discussions are clearly justified by the numerical simulations as shown in Fig. 3.

The wave functions $\vert \psi_w\rangle$ and  $\vert \psi_p\rangle$  describe  wavelike and particlelike behavior, respectively.  When the system is finally prepared in $\vert \psi_w\rangle$, then the probability for detecting the spin in   state $\vert e\rangle$ is $\cos^2(\phi/2)$, displaying $\phi$-dependent interference patten associated with waves.
On the other hand, if the system is finally prepared in $\vert \psi_p\rangle$, then the probability for detecting the spin in   state $\vert e\rangle$ or $\vert g\rangle$ is $1/2$.
In this case, the interference patten disappears and it represents particlelike behavior.
The visibility of the interference patten is $\mathcal {V}=(P_\text{max}-P_\text{min})/(P_\text{max}+P_\text{min})$, where the min/max values are calculated with
respect to $\phi$ for both the wave and particle cases. For the case of the spin described by $\vert \psi_w\rangle$, we have $\mathcal {V}=1$, while for the
case of $\vert \psi_p\rangle$, we get $\mathcal {V}=0$.
The transformations $R^1_{\pi/2}$ and $R^2_{\pi/2}$ can be considered as beam splitters for
spin states in the Hilbert space,  similar to the beam splitters  for photons in the Mach-Zehnder interferometer.

We can investigate the wavelike and particlelike aspects of the NV spin simultaneously by preparing the mechanical resonator in a quantum superposition state,
i.e., $\cos \alpha\vert 0\rangle+\sin\alpha\vert N_0\rangle$. Then, after the two rotations $R^1_{\pi/2}$ and $R^2_{\pi/2}$,  the final state becomes
\begin{eqnarray}
 \vert \psi\rangle&=&\cos\alpha\vert \psi_p\rangle\vert 0\rangle+\sin\alpha\vert \psi_w\rangle\vert N_0\rangle.
\end{eqnarray}
The probability for detecting the spin in  state $\vert e\rangle$ thus becomes
\begin{eqnarray}
P_e=\frac{1}{2}\cos^2\alpha+\sin^2\alpha\cos^2\frac{\phi}{2},
\end{eqnarray}
with the corresponding visibility $\mathcal {V}=\sin^2\alpha$.
By varying the parameter $\alpha$, we have the ability to modify continuously the Ramsey interference patten. Fig. 4 shows the probability $P_e$ as a function of
$\alpha$ and $\phi$, which demonstrates a morphing behavior between a particle ($\alpha=0$) and a wave  ($\alpha=\pi/2$). In the case of $\alpha=\pi/4$, the spin in state $ \vert \psi\rangle$ exhibits both wavelike and particlelike behavior with equal probability. Therefore, we can measure both properties in a single experiment without the need to change the experimental setup in order to test Bohr's complementarity principle.

\begin{figure}[t]
\centerline{\includegraphics[bb=89 270 519 564,totalheight=2.3in,clip]{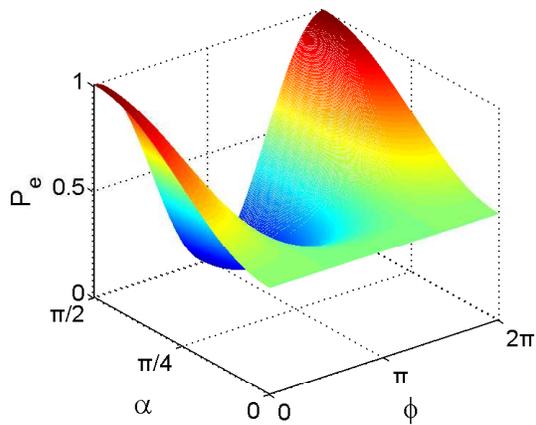}}
\caption{(Color online) Morphing behavior between particle ($\alpha=0$) and wave ($\alpha=\pi/2$) of the probability for detecting the spin in state $\vert e\rangle$. }
\end{figure}

\section{Practical considerations and experimental parameters}
In realistic situations, we need to
consider spin dephasing (with a decay rate $\gamma_s$) and mechanical dissipation (with a decay  rate $\gamma_m$). The full dynamics of this
system that takes these incoherent processes into account
is described by the following  master equation
\begin{eqnarray}
\label{M1}
\frac{d\hat{\rho}(t)}{dt}&=&-i[\hat{\mathcal{H}},\hat{\rho}]+\gamma_s\mathcal{D}[\vert e\rangle\langle e\vert]\hat{\rho}\nonumber\\
&&+n_\text{th}\gamma_m\mathcal{D}[\hat{a}^\dag]\hat{\rho}+(n_\text{th}+1)\gamma_m\mathcal{D}[\hat{a}]\hat{\rho},
\end{eqnarray}
where
\begin{eqnarray}
\hat{\mathcal{H}}&=&[\Omega_1(t)+\Omega_2(t)](\vert e\rangle\langle g\vert+\vert g\rangle\langle e\vert )+\Delta(t)\vert e\rangle\langle e\vert+\hat{\mathcal{H}}_s,\nonumber\\
\end{eqnarray}
and $\mathcal{D}[\hat{o}]\hat{\rho}=\hat{o}\hat{\rho}\hat{o}^\dag-\frac{1}{2}\hat{o}^\dag\hat{o}\hat{\rho}-\frac{1}{2}\hat{\rho}\hat{o}^\dag\hat{o}$
for a given operator $\hat{o}$. In addition, $n_\text{th}=(e^{\hbar \omega_m/k_\text{B}\mathcal {T}}-1)^{-1}$ is the thermal phonon number at the environment
temperature $\mathcal {T}$.
We need to note that the decoherence model is reasonable in this spin-mechanical setup and it is fair and unbiased for checking the feasibility of this protocol.
In Fig. 5, we display the numerical simulations of quantum dynamics of the driven spin-mechanical system through
solving the master equation (\ref{M1}). To perform the simulations, we choose the following sequence of pulses for simplicity
\begin{eqnarray}
\Omega_1(t)&=&\left\{
                \begin{array}{ll}
                  \Omega_1, &  0\leq t\leq\pi/4\Omega_1 \\
                  0, &t>\pi/4\Omega_1,
                \end{array}
              \right.\nonumber\\
\Delta(t)&=&\left\{
              \begin{array}{ll}
              0, &0\leq t<  \pi/4\Omega_1 \\
              \Delta,   &\pi/4\Omega_1\leq t \leq\pi/4\Omega_1+ T      \nonumber\\
              0, & t>\pi/4\Omega_1+ T,\\
              \end{array}
            \right.\\
\Omega_2(t)&=&\left\{
                \begin{array}{ll}
                0, & 0\leq t<\pi/4\Omega_1+T\\
                \Omega_2, &  \pi/4\Omega_1+T\leq t\leq\pi/4\Omega_1+T+\pi/4\Omega_2  \nonumber\\
                0, &t>\pi/4\Omega_1+T+\pi/4\Omega_2.
                \end{array}
              \right.
\end{eqnarray}
\begin{figure}[t]
\centerline{\includegraphics[bb=19 132 260 397,totalheight=3.2in,clip]{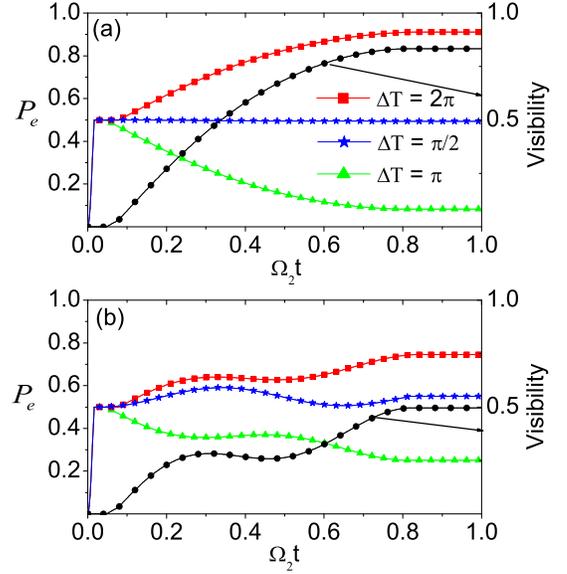}}
\caption{(Color online) Numerical simulations for the system dynamics and the visibility of the interference pattern with two different initial states, (a) $\vert g,1\rangle$ and (b) $\frac{1}{\sqrt{2}}(\vert g, 0\rangle+\vert g,1\rangle)$. The relevant parameters are $\Omega_1=50\Omega_2$, $\chi=10\Omega_2$, $\Delta=100\Omega_2,\gamma_s=0.1\Omega_2$,
and $n_\text{th}\gamma_m=0.1\Omega_2$.}
\end{figure}

Fig. 5 (a) displays that the system starts from the state $\vert g,1\rangle$ and evolves into the state $\vert \psi_w\rangle$ with several phase differences $\phi=\Delta T$. At the end of the Ramsey rotations, we find that the probabilities for detecting the spin in state $\vert e\rangle$ are $P_e=0.92,0.49$ and $0.08$, corresponding to $\phi=2\pi,\pi/2$, and $\pi$, respectively, while the corresponding visibility is $\mathcal {V}=0.84$.
Fig. 5 (b) presents the results for the case where the initial state is $\frac{1}{\sqrt{2}}(\vert g, 0\rangle+\vert g,1\rangle)$.
At the end of the Ramsey rotations, we find $P_e=0.74,0.54,$ and $0.25$, respectively, in  agreement with the analytic expression $P_e=\frac{1}{2}\cos^2\alpha+\sin^2\alpha\cos^2\frac{\phi}{2}$. Moreover, at the end of the process, the corresponding visibility is $\mathcal {V}=0.495$, in good agreement with the expression $\mathcal {V}=\sin^2\alpha$ when $\alpha=\pi/4$.
Thus, this proposal works very well under realistic conditions.

We now proceed to consider the experimental feasibility of this proposal and the appropriate
parameters. An NV center in diamond consists of a substitutional
nitrogen atom and an adjacent vacancy \cite{PR-528-1}, which has a spin $S=1$
ground state, with zero-field splitting $D=2\pi\times 2.87$ GHz,
between the $\vert m_s=\pm1\rangle$ and $\vert m_s=0\rangle$ states. For moderate applied magnetic fields,  one of the spin transitions  can be tuned into near resonance with the mechanical mode and external microwaves. Here we can choose $\vert g\rangle =\vert m_s=0\rangle$, and $\vert e\rangle =\vert m_s=1\rangle$.
Through the action of an external magnetic gradient \cite{prb-79-041302}, NV centers can interact with nanomechanical resonators in the dispersive regime.
We can choose $\chi/2\pi\sim 100$ kHz, $\Omega_2/2\pi\sim 10$ kHz, $\Omega_1/2\pi\sim 500$ kHz,  $\Delta/2\pi\sim 1$ MHz, and $T\sim 1 /\Delta$.
All these parameters are in line with current experimental techniques. Then,
the required time for completing all the processes is about 10 $\mu$s, which is much shorter than the time scales associated with spin dephasing \cite{Naure-Mat} and mechanical
dissipation \cite{Natcomm-5-3638}.

We need to point out that  quantum delayed-choice experiments have been performed with optical \cite{SCI-338-634,SCI-338-637,nphton-6-600} systems using
MZ interferometers. These experiments are on the microscopic level, i.e., they examine the wave or particle properties of single photons, while our protocol is on the macroscopic scale, i.e., we examine the wave or particle properties in a spin-mechanical setup. Different from the optical experiments based on MZ interferometers,  this scheme does not need the spacelike separation as required in Wheeler's original  thought experiment, due to the use of a temporally based Ramsey interferometer. Here the beam splitters are for spin states in the Hilbert space, while in \cite{SCI-338-634,SCI-338-637,nphton-6-600} the beam splitters in the MZ interferometer are for single photons in real space.

Compared to the work \cite{prl-115-260403}, this protocol has  two different features: (i)  In this double-pulse Ramsey interferometer, the second
beam splitter is based on the selective spin rotations conditional on the number states of
the mechanical resonator. This results from a dispersive coupling between the NV spin and the mechanical resonator, which produces a selective drive in the spin-motion Hilbert space. In Ref. \cite{prl-115-260403}, however, the first beam splitter is produced by the resonant interaction between the qubit and the cavity
mode, which is either in a coherent microwave photon state or  its photonic vacuum state. (ii) In both protocols, a superposition state of the resonator mode is required, in order to prepare  one of the beam splitters  in a superposition of being active and inactive. In our work, only a superposition of two number states is needed, which is quite easy in the experiment. However, in Ref. \cite{prl-115-260403} they need to prepare the resonator in the cat superposition state, i.e., a superposition of a coherent state and the vacuum state. As pointed out in Ref. \cite{prl-115-260403}, the coherent state $\vert \alpha\rangle$ is not strictly orthogonal to $\vert 0\rangle$, so that these two components cannot be unambiguously discriminated. Furthermore, the realization of a cat state is more difficult than the
realization of a superposition of two number states of cavity modes in the experiment.

\section{Conclusions}
In conclusion, we have proposed an experimentally feasible scheme for implementing a quantum delayed-choice experiment with the form of
Ramsey interferometer in a spin-mechanical setup.  In this proposal, a single NV center in diamond driven by
microwave fields is dispersively coupled to a massive mechanical resonator, enabling  selective  spin rotations
conditional on the number states of the resonator.  The probability
for finding the NV center in  the upper spin state exhibits interference patten when the mechanical
resonator is prepared in a specific number state. On the other hand, the interference is destroyed
if the mechanical resonator stays in some other number states. The wavelike and particlelike aspects of the NV spin can be simultaneously
investigated by preparing the
mechanical resonator in a quantum superposition  of its number states. In this sense, a quantum version of Wheeler's delayed-choice experiment can be
implemented, allowing us to test quantum mechanics with massive objects.

\section*{ACKNOWLEDGMENTS}
This work is supported by the NSFC under Grants
No. 11474227 and No. 11534008.   Part of the
simulations are coded in PYTHON using the QUTIP library \cite{CPC}.

%

\end{document}